\documentclass{PoS}

\title{How do diquark fluctuations and chiral soft modes 
affect di-lepton production in the deconfined phase?}

\ShortTitle{di-lepton production due to diquark and chiral fluctuations}

\author{\speaker{Teiji Kunihiro}
\\
        Yukawa Institute for Theoretical Physics,
Kyoto University, Kyoto 606-8502, Japan\\
        E-mail: \email{kunihiro@yukawa.kyoto-u.ac.jp}
}

\author{Masakiyo Kitazawa\\
       Department of Physics, Osaka University,
Toyonaka, Osaka 560-0043, Japan\\
       E-mail: \email{masky@yukawa.kyoto-u.ac.jp}
}

\author{Yukio Nemoto\\
      Department of Physics, Nagoya University,
Nagoya 464-8602, Japan\\
      E-mail: \email{nemoto@hken.phys.nagoya-u.ac.jp}
}

\abstract{
We examine diquark fluctuations and chiral soft modes existing in the
precritical region of  color superconductivity and chiral transition, respectively, at
finite temperature and density. 
We evaluate how they contribute to anomalous di-lepton production; 
although there appear peaks in the spectral function owing to the
existence of the soft modes, the enhancement of the production rate 
may not be so prominent to be a clear signal of the phsase transitions.
}

\FullConference{Critical Point and Onset of Deconfinement
          4th International Workshop\\
		 July 9-13 2007\\
		 GSI Darmstadt,Germany}

\begin{document}

\section{Introduction}

In this report, we focus on the precursory phenomena of
two of QCD phase transition at finite temperature and/or density,
i.e., the chiral transition and the color superconductivity.
We assume that the phase transitions are almost of second order
and hence specific soft modes exist around the respective critical
points.

\section{Precursory phenomena of color superconductivity in heated
quark matter}

In this part, we shall consider quark matter 
{\em in the normal phase} but near the critical temperature $T_c$ 
of the color superconductivity and discuss the possibility 
to observe pre-critical phenomena of
color superconductivity by the heavy-ion collisions.

\begin{figure}
\includegraphics[scale=0.4,angle=0]{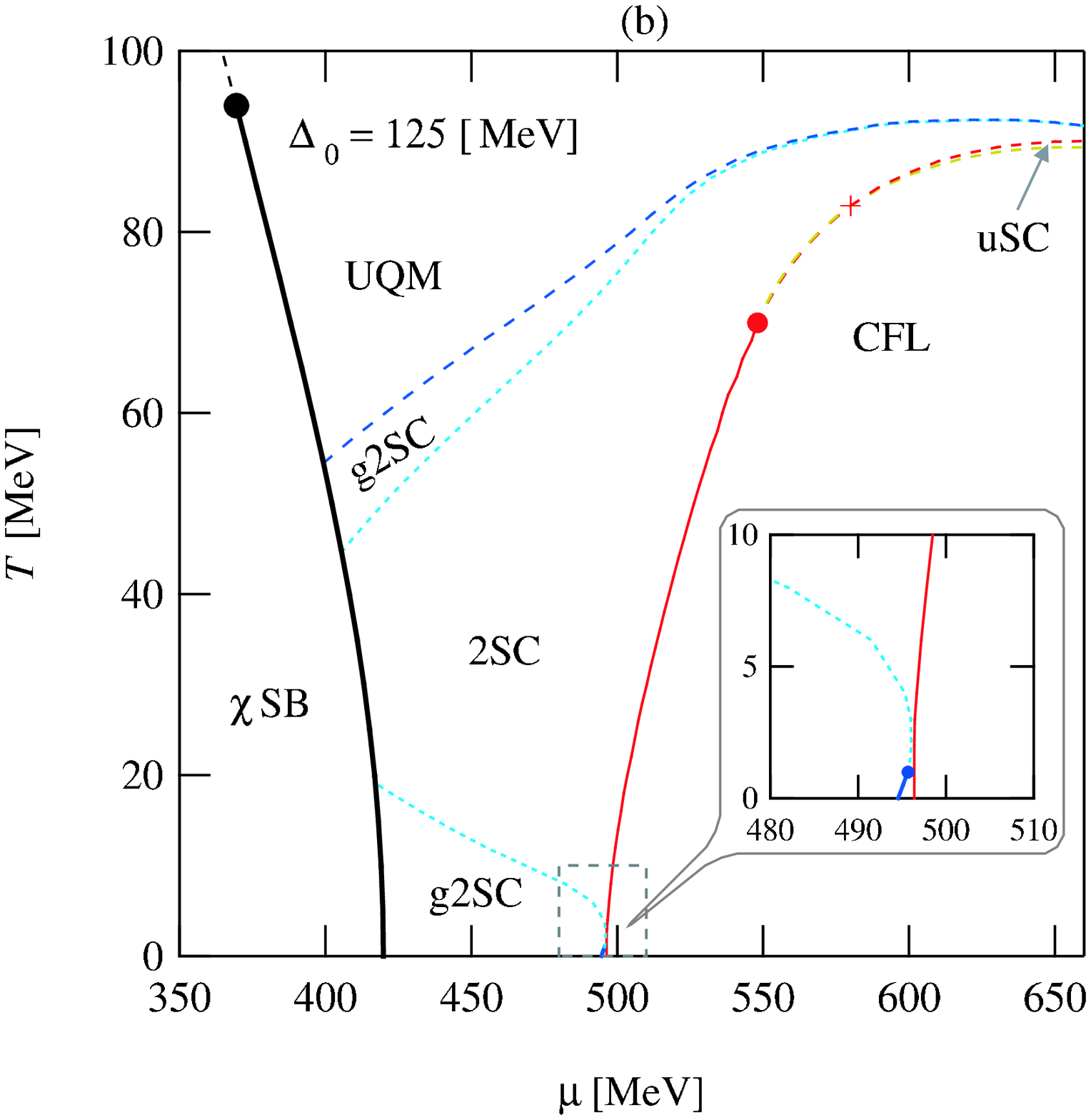}
\includegraphics[scale=0.4,angle=0]{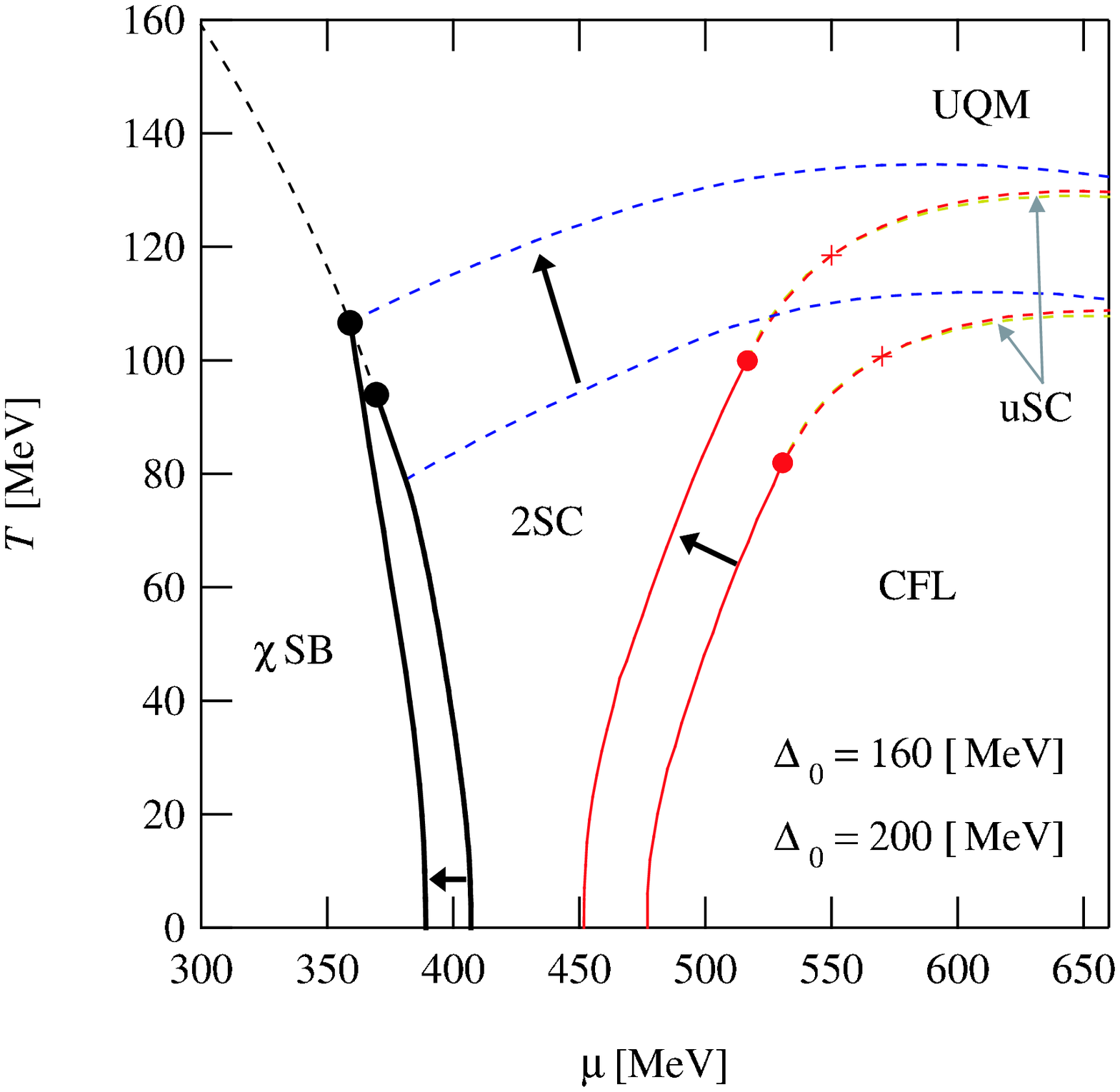}
\caption{Left panel; the phase diagram for the intermediate
strength of the pairing interaction for the three flavor (q$=$u, d, s)
case under the charge and color neutrality constraint.
Right panel; the phase diagram for the large strength of the 
pairing interaction. Others are the same as the left panel.
Taken from \cite{Abuki:2005ms}; see this reference for the details.}
\label{phased-abuki}
\end{figure}

Owing to the many internal degrees of freedom of quarks,
there are rich varieties of
 the pairing patterns of the color superconductivity.
However, the highest-$T$ phase which
is  relevant for heavy-ion collisions at high baryon densities
 is the two-flavor superconducting
(2SC) phase\cite{Abuki:2005ms}.
We show examples for the 2SC phase boundaries obtained
in Nambu-Jona-Lasinio-type models in 
Fig. \ref{phased-abuki}.

\begin{figure}
\begin{center}
\includegraphics[scale=0.4, angle=0]{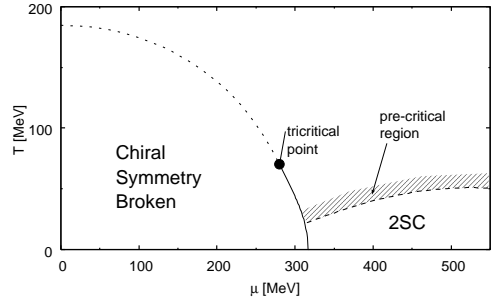}
\caption{
Phase diagram for isospin-symmetric two-flavor quark matter 
from an NJL-type model with a relatively weak 
coupling \cite{Kitazawa:2005vr}. The critical temperatures for the 
2SC phase transition is model-dependent, 
as seen from the comparison with Fig. 1.}
\label{phased-kitazawa}
\end{center}
\end{figure}

We explore the possibility to 
see precursory phenomena of color superconductivity
in the quark matter possibly created by the heavy-ion collisions. 
Since we are interested in  the high-$T$ phase because heavy-ion collisions
can only create heated matter, we may thus focus on the 2SC phase.
Our discussions are based on the observation that 
there can exist a rather wide 
pre-critical region of the color superconductivity 
in the $T$-$\mu$ plane at moderate density\cite{Kitazawa:2001ft},
as shown in Fig.~\ref{phased-kitazawa}, where the small pairing
coupling is taken, and the neutrality constraints are not imposed.
The large diquark-pair fluctuations may affect various observables
 leading to  precursory phenomena\cite{Kitazawa:2001ft,Kitazawa:2003cs}. 

In the extremely high-density region,
the perturbative calculation should be valid
and the phase transition to the color-superconducting phase is 
a first order due to the gauge fluctuations in this region.
On the other hand, the color superconductivity is expected 
to turn to a type-II at lower density\cite{Giannakis:2003am}.
We shall only consider the effects of the pair fluctuations at moderate density 
which is relevant to heavy-ion collisions.

We first notice that 
the ratio of the diquark coherence length to the average 
inter-quark length, which is proportional to $E_F/\Delta$
with $E_F$ being the Fermi energy, can be as small as $\sim 10$
at moderate densities.
The stronger the  interaction between the quarks,
the shorter the  coherence length, which can be
as small as almost the same order of the inter-quark 
distance\cite{Matsuzaki:1999ww,Abuki:2001be}.
The short coherence length implies that the fluctuation of 
the pair field is significant and the mean-field approximation
looses its validity.

\begin{figure}[tb]
\begin{center}
\includegraphics[scale=.6]{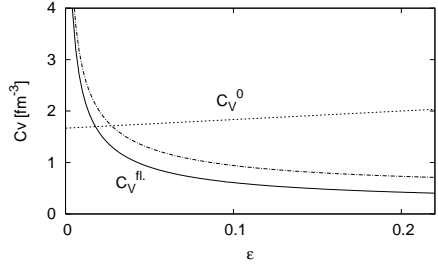} 
\caption{
The specific heat per unit volume 
$c_{\rm v}^{\rm fl.}$ 
with the effects of the fluctuations of the pair field,
together with that in the free fermionic system, $c_{\rm v}^0$.
The total specific heat is  given by
$c_{\rm v} = c_{\rm v}^0 + c_{\rm v}^{\rm fl.}$.
The dot-dashed curve represents the specific heat obtained with 
only the static part of $\Omega_{\rm fl.}$.
$\varepsilon\equiv (T-T_c)/T_c$.}
\label{fig:SH}
\end{center} 
\end{figure}

The large fluctuations cause an excess 
of the specific heat, which eventually diverges at $T_c$
owing to the critical fluctuations for the second order
 transition\cite{Kitazawa:2005vr}, as shown in Fig.\ref{fig:SH}:
We can see that as the temperature  approaches $T_c$ from
above,  the anomalous part of the specific heat $c_{\rm v}^{\rm fl.}$ become
 significant.
Such an anomalous increase of the specific heat may
affect the cooling of the proto-compact stars.
Here one should, however, notice that
the critical divergence of the specific heat is essentially 
due to the static fluctuations of the pair field\cite{LL58};
the temperature region where  static fluctuations is significant 
\cite{Voskresensky:2004jp,Kitazawa:2005vr}.

We now turn to the discussions on the dynamical
fluctuations of the diquark pair field.
At finite temperature, 
the dynamical fluctuations of the  pair field become also 
significant and develop a well-defined collective mode 
as the temperature is lowered toward $T_c$
if the color superconducting phase transition is of second order\cite{Kitazawa:2001ft}:
 The spectral function
of the diquark fluctuations gets to have a sharp peak 
in the low-energy region at about $T = 1.2T_c$, and 
the peak position decreases as the temperature is lowered 
toward the critical temperature.
This collective soft mode is found to be 
a diffusive mode.

\begin{figure}
\begin{center}
\includegraphics[scale=0.7,angle=0]{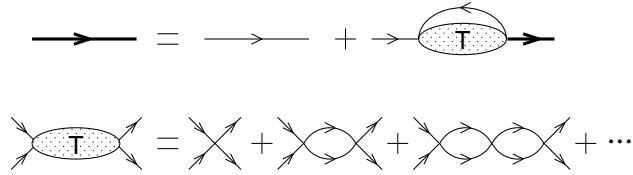}
\caption{
The Feynman diagrams representing the
quark Green function in the T-matrix approximation
employed in Ref.~\cite{Kitazawa:2005vr}.
The thin (bold) lines represent the free (full) propagator
and the wavy line denotes the pairing soft mode.
}
\label{fig:T-matrix}
\end{center}
\end{figure}

The soft mode of the color superconductivity
in the heated quark matter
leads to  the formation of 
a pseudogap in the density of states (DOS) of quarks, i.e.,
an anomalous depression in the DOS
around the Fermi surface\cite{Kitazawa:2003cs,Kitazawa:2005vr};
the quark propagator is modified due to 
the coupling with the fluctuating pair-field or the pairing soft mode,
as shown in Fig.~\ref{fig:T-matrix}.

\begin{figure}
\begin{center}
\includegraphics[scale=0.6,angle=0]{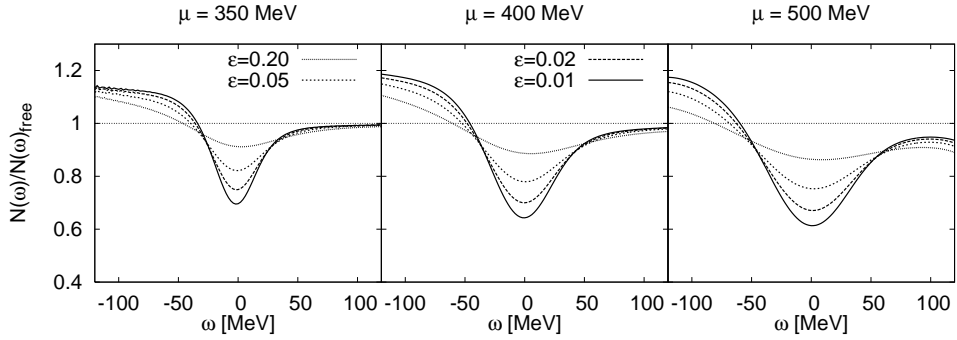}
\caption{
Density of states for quark matter near but {\em above} $T_c$
with several reduced temperatures $\varepsilon$ 
and quark chemical potentials $\mu=350,400,500$MeV\cite{Kitazawa:2003cs}.
One sees a clear pseudogap phenomenon irrespective of $\mu$
near $T_c$.}
\label{fig:DOS}
\end{center}
\end{figure}

As shown in Fig.~\ref{fig:DOS},
there appears a depression in the DOS around the Fermi energy
for each $\mu$ near $T_c$,
and they survive up to $\varepsilon  \equiv ( T-T_c )/T_c \approx 0.1$
irrespective of $\mu$\cite{Kitazawa:2003cs,Kitazawa:2005vr}.
This is an incomplete gap, or  ``pseudogap''
formed within the QGP phase above $T_c$.
The pseudogap in the DOS
implies  that the quarks around the Fermi surface
have a short life time owing to the decay process 
q$\rightarrow \,$ hole$+($qq$)_{\rm soft}$ emitting 
the soft mode like Cherenkov process.
This short-livedness of the quarks around the Fermi surface
means that the system 
is {\em not} a Fermi liquid
\cite{Kitazawa:2005vr,Kitazawa:2005pp}.
We notice  that the pseudogap formation is known 
as a characteristic behavior of the materials which become
the high-$T_c$ superconductors(HTSC)\cite{TS99,HTSC}. 
Thus we can say that the heated quark mater at moderate densities
is similar to
the HTSC materials  rather than the usual superconductors of metals.

\begin{figure}
\begin{center}
\includegraphics[scale=0.5]{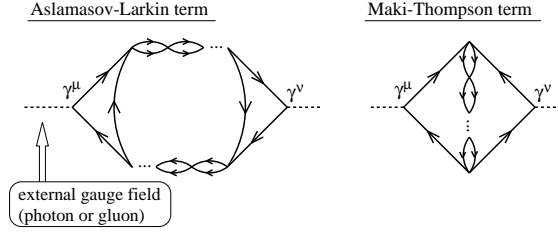}
\caption{
The diagrams that contribute to the photon (or gluon) self-energy 
representing the Aslamazov-Larkin (left) and the Maki-Thompson (right)
terms\cite{Kitazawa:2005vr}.
The wavy lines denote the soft mode.}
\label{fig:AL}
\end{center}
\end{figure}

It is known 
that pair fluctuations above $T_c$ cause
a large excess of the electric conductivity, which is  
called the paraconductivity in condensed matter physics. 
Two microscopic mechanisms that 
give rise to such an anomalous  conductivity
are identified in terms of Feynman diagrams.
They are called Aslamazov-Larkin (AL) and 
Maki-Thompson terms\cite{ref:AL-MT}, 
both of which are depicted in Fig.~\ref{fig:AL};
the dotted lines in the figure denote the gauge field, i.e.,
 the photon in this case.
The {\em color-conductivity} would be also enhanced
by the similar mechanisms near $T_c$,
although  it would be of academic interest only.
Our point is that the {\em photon} self-energy $\Pi^{\mu\nu}(Q)$
in the quark matter at $T>T_c$ can be 
also modified due to the fluctuations of the diquark pair-field
as well as inside the color superconducting phase\cite{jai:2001}:
The diagrams shown in Fig.~\ref{fig:AL}
can be interpreted as modifications of the self-energy of the gauge
fields, i.e., $\Pi^{\mu \nu}(Q)
={\cal F}[i\theta(t)\langle[j^{\mu}(x),j^{\nu}(0)\rangle]$,
where ${\cal F}$ denotes the Fourier transformation.
The external photon field thus can couple to 
the soft mode of the color superconductivity
with the pairing soft mode in the diagrams
 being replaced by the diquark pair fields.
This is interesting, because
modifications of the photon self-energy
 may be detected as an enhancement of 
the invariant-mass distribution of 
the dileptons emitted from the created matter.
The lepton-pair production rate per four-momenta
is given by the 
well-known formula\cite{LeBellac}
\begin{eqnarray}
\frac{ d\Gamma }{ d^4 Q } 
= \frac{ -\alpha g^{\mu\nu} {\rm Im} \Pi_{\mu\nu}(Q) }
{ 12 \pi^4  Q^2 ( {\rm e}^{\beta q_0} - 1 ) }.
\label{eq:drate-0}
\end{eqnarray}
The more direct observable in experiment
is  the invariant mass spectrum of dilepton production rate, which is given
by integrating  $ d\Gamma/\,d^4 Q$ over four momenta
\begin{eqnarray}
\frac{ d\Gamma }{dM^2}
= \int \frac{ d^3 q }{ 2q^0 } \frac{ d\Gamma }{ d^4 Q }.
\label{eq:drate_precursor}
\end{eqnarray}

\begin{figure}
\begin{center}
\includegraphics[scale=0.8]{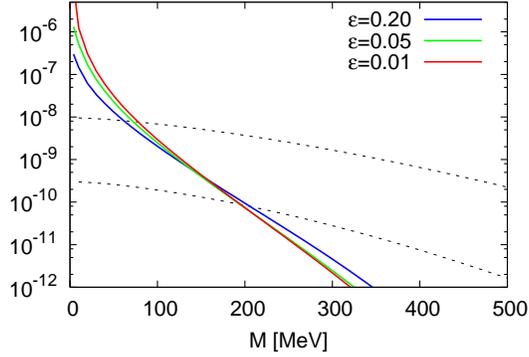}
\caption{
A preliminary result for the dilepton production rates 
Eq.(2.2) for several reduced temperatures 
$\varepsilon \equiv ( T-T_c )/T_c$  near but above $T_c$ 
with $\mu=400$MeV \cite{KKprep}. The solid lines include 
the contributions of the Aslamazov-Larkin term,
while the dashed lines represent the dilepton rate from the 
free quark matter  at $T=T_c$ (below) and $T=1.5T_c$ (above).
}
\label{fig:drate_precursor}
\end{center}
\end{figure}

We show in Fig.~\ref{fig:drate_precursor} 
a preliminary result\cite{KKprep} of  $d\Gamma/ dM^2$ obtained from
the $\Pi_{\mu\nu}(Q)$ with the Al term, which is evaluated
in some approximation\cite{KKprep}:
The solid lines denote the production rate with
the AL term  for some reduced temperatures
$\varepsilon \equiv ( T-T_c )/T_c$ at $\mu=400$MeV, 
while the dashed lines show the production rate 
from the free quark system
for $T=T_c$ and $1.5T_c$ at $\mu=400$MeV.
The Figure shows that the contribution of the AL term causes
a large enhancement with a sharp-peak structure in the production rate 
in the lower energy region,  and the peak becomes 
larger and sharper as the temperature approaches $T_c$.
It is known that a similar behavior of $ d\Gamma / dM^2 $
is also seen in the color superconducting phase itself\cite{jai:2001}.
In both cases, the characteristic enhancement of $ d\Gamma / dM^2 $
appears at energies less than the pion mass.
Therefore, experimental observations of the dilepton production 
rate at low energy might be able to detect an signature of
a formation or a precursor of the color superconductivity
in the heavy-ion collisions.
Such a low-mass energy region is, however, also the region where
other hadronic and electro-magnetic process contribute to the
dilepton production. So a dis-entanglement from these processes
have to be made to identify the AL process due to the precursory
diquark pairing  fluctuations, which might be, unfortunately, difficult
to perform.
One should  estimate also the contribution from the
Maki-Tompson term as well as  confirm the present estimate
of the AL term more definitely.

\section{The soft modes of chiral transition and 
enhancement of lepton-pair production}

The chiral transition is a QCD phase transition with the order
parameter $\langle \bar{q}q\rangle$.
If the phase transition is of second order, of weak first order or
of cross over, there should be a critical region where the chiral
fluctuations $\langle (\bar{q}q)^2\rangle$ and 
$\langle (\bar{q}i\gamma_5\tau_a q)^2 \rangle$ in the scalar and pseudoscalar
channels, respectively, significant.
It means that there will exist elementary soft modes or quasi-particles
corresponding to these fluctuations in the critical region\cite{HK84}.
In the following discussions, we will assume that the chiral transition
at finite temperature and/or density has a critical region where
such chiral soft modes have significant strength.
In this context, it is noteworthy that the lattice people determine
the critical temperature of the chiral transition from the peak 
position of the chiral susceptibility in the scalar channel, i.e.,
$\chi_m\equiv \partial /\partial m\langle\langle \bar{q}q\rangle\rangle=
\langle\langle (\bar{q}q)^2\rangle\rangle$; which surely shows a
peak behavior around some temperature\cite{karsch02}.
In other words, the so called generalized mass squared defined by
$m_{\sigma}^2\equiv \chi_m^{-1}$ decreases as $T$ goes high
and gives the minimum around the critical
temperature and then increases in the chirally symmetric phase;
in the high $T$ phase, the generalized masses in the $\sigma$ and 
the $\pi$ channels tend to be degenerated.

Recently, people are much interested in the properties of the QGP phase 
near the critical temperature ($T_c$)\cite{RHIC}:
The success of the perfect hydrodynamics in 
reproducing  the elliptic flow of hadrons in the
heavy-ion collisions at the Relativistic Heavy Ion Collider (RHIC)
suggests that the created matter is a strongly coupled system.
One of the interesting problems on the QGP phase near $T_c$ is
the possible existence of hadronic excitations  even in the
QGP phase near $T_c$ \cite{HK84,Lattice}.
The existence of the mesonic excitations in the light-quark sector 
was suggested  as being the soft modes associated to 
the chiral transition \cite{HK84}:
The spectral function calculated in the NJL model
shows the existence of degenerated hadronic soft modes called 'para-pion' and
'para-sigma' at $T>T_c$.
It was found that the soft modes acquire a large strength but with
small width as $T$ approaches $T_c$ from above; it means that 
the soft modes become a good elementary modes in the vicinity of
$T_c$.

\subsection{QCD phase transition at finite $T$ and  $\mu$ with finite quark mass}

In the chiral limit with two flavors (q$=$u, d), 
the chiral transition at finite temperature with vanishing
chemical potential $\mu$ is known to be second order\cite{asakawa-yazaki}:
As $\mu$ is raised,
the critical line of the second-order transition extends up to
some critical point 
P$_{{\bf TCP}}$:$(T_{{\bf TCP}}, \mu_{{\bf TCP}})$ in the $(T, \mu)$-plane.
The point P$_{{\bf TCP}}$ is  a tri-critical point because from it
the critical line turns to a critical line of the first-order transition
for higher density and lower temperatures.
When the current quark masses are introduced, the second-order transition
at small $\mu$ region changes to a cross-over\cite{asakawa-yazaki,TK89}.
 The common belief
based on calculations of some chiral effective models\cite{stephanov} is  that 
the critical line of the first-order transition in the large $\mu$
region remains; the tricritical point changes its nature to a critical end 
point P$_{{\bf CEP}}$.

\begin{figure}
\begin{center}
\includegraphics[scale=1]{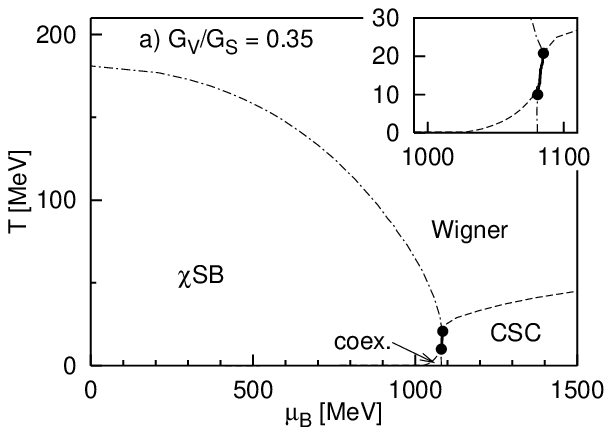}
\includegraphics[scale=1]{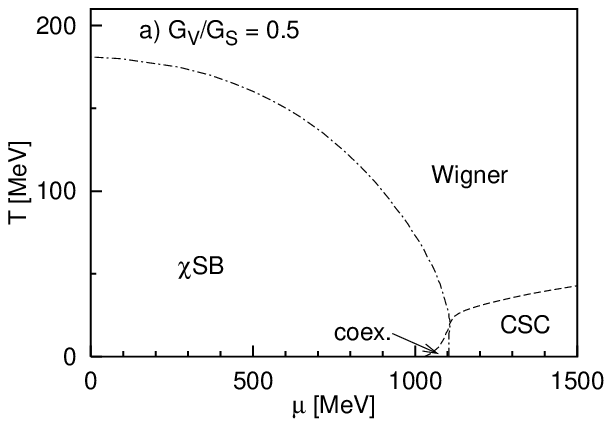}
\caption{
The left panel: The phase diagram with $G_V/G_S=0.35$
in the $T$-$\mu$ plane.
There appear two endpoints of the first-order transition.
The right panel: The phase diagram 
in the $T$-$\mu$ plane with $G_V/V_S=0.5$.    
The critical line for the first-order transition
disappears completely and the 
critical end point disappears with the large vector coupling.    }
\label{fig:vector-coupling}
\end{center}
\end{figure}

It should be noticed, however,
that  there is some caveats to this belief that the QCD phase transition
at small temperature is of first order.
It has been known for some time \cite{asakawa-yazaki}
that the inclusion of the vector term 
$G_{_V}(\bar{q}\gamma_{\mu}q)^2$ can drastically alter the nature
of the chiral transition at small temperature but at finite density
where the color superconductivity was not taken into account;
with an increase of the coupling constant $G_{_V}$, the position of
P$_{{\bf CEP}}$ moves to lower $T$ and higher $\mu$, and eventually disappear
in the $(T, \mu)$-plane! When the color superconductivity is incorporated,
more interesting phenomenon can occur\cite{kitazawa02} as shown 
in Fig. \ref{fig:vector-coupling}:
It is noteworthy that 
there appear another critical end point in the lower
$T$, and hence there can exist {\bf two end points} 
at both sides of the critical line 
of the first-order transition for the vector coupling 
$G_{_V}/G_{_S}=0.35$ where the $G_{_S}$ denotes the strength of
the scalar coupling $(\bar{q}q)^2$ in the NJL model.
We find that  the coexisting-CSC transition at low temperatures 
becomes a crossover transition, and there exists a coexisting phase
where both the chiral and the diquark condensates have finite values.
For larger $G_{_V}$, the critical line for the first-order transition
disappears completely and there exists only the cross-over in the
entire $(T, \mu)$ plane.
It is interesting that such a coexisting phase of the chirally-broken and
color superconducting phase and hence the appearance of the new
critical end point has been advocated where the driving force
of the coexistence is not due to the vector coupling but the
axial anomaly term\cite{Hatsuda-anomaly}.

\subsection{The soft modes around the critical end point}

The phase transition at the critical end point P$_{{\bf CEP}}$ is of second order.
Then what is the soft mode for this second order transition?
The $\sigma$ meson as the fluctuation of the chiral condensate 
has still a non-zero mass at P$_{{\bf CEP}}$ because the chiral symmetry is
explicitly broken. 
Fujii\cite{fujii} showed that the soft mode is a hydrodynamical mode which
has the strength in the space-like region of the momentum-energy plane:
At finite density, the charge conjugation symmetry is broken and
the scalar-vector mixing can occur and the soft mode in this case
consists of the density fluctuation and the chiral fluctuation, of which
the former is the main component of the mode.
Since this soft mode exists in the space-like region, it can not affect
particle productions to be seen in the time-like region.
It can, however, contribute to the hydrodynamical modes, leading to 
dynamical critical phenomena\cite{son-stephanov}.

\subsection{Di-lepton production owing to the scalar-vector mixing}

Although the sigma-mesonic  mode is not a genuine soft mode
around P$_{{\bf CEP}}$, it can show a slight softening as the system 
approaches the critical temperature from above.
So we shall examine  effects of the sigma mesonic mode in the time-like
region on the lepton-pair production\cite{nemoto}.
The relevance of the scalar mode to the photon properties in the
system can be understood as follows.
As we mentioned in the previous subsection, 
there arises the scalar-vector mixing at finite $\mu$;
the zero-th component of the vector is relevant and it is the 
baryon density.
The importance of such a mixing was  noticed in the work 
on the baryon-number susceptibility in association of the
chiral transition\cite{baryon}; it was first shown that
 the quark-number susceptibility can be enhanced around the critical
point of the chiral transition at finite density as well as finite temperature.

The baryon-number susceptibility is a static quantity.
The scalar-vector mixing also occurs in the dynamical cases.
Owing to the mixing, the photon self-energy can be affected
by the possible change of the $\sigma$ mode mass.

We calculate the invariant mass distribution of electron pairs
using the formula given in Eq.{\ref{eq:drate-0}) where the photon
self-energy include the contribution from the soft mode
as given in Fig.\ref{fig:fluc-loop} as well as the 
bare one given in Fig.\ref{fig:Feynman-di-lepton}.
The resultant electron-pair production rate
 is shown in the left panel of Fig\ref{fig:di-lepton}, 
where the temperature dependence of the production rate
is shown; The upper curves correspond to higher temperature
away from $T_c$.
We can see an enhancement of the production, which is encouraging but
unfortunately may not be sufficient for detecting in experiments.
Here it should be noted that the figure shows the invariant mass
distribution of the lepton pair where the out-coming momenta are
integrated out. However, the momentum-non-integrated probability
show a clear peak, although not shown here.
 
\begin{figure}
\begin{center}
\includegraphics[scale=1]{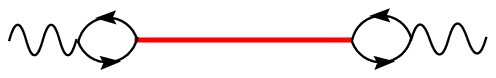}
\includegraphics[scale=.8]{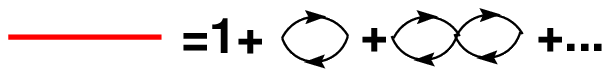}
\caption{
The left graph shows the anomalous photon self-energy owing to the
chiral soft mode denoted by the straight line in the middle.
The straight line is actually composed of the scalar bubble diagrams as
shown in the right panel. }
\label{fig:fluc-loop}
\end{center}
\end{figure}

\begin{figure}
\begin{center}
\includegraphics[scale=.4]{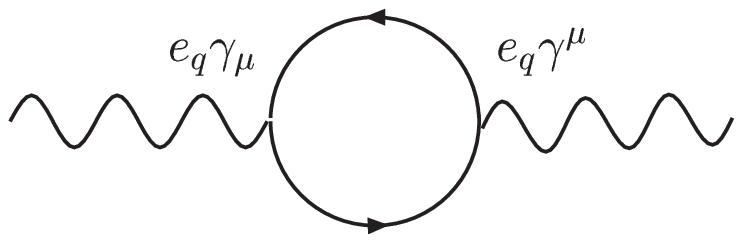}
\caption{The bare self-energy diagram of the photon due to the quark loop.
}
\label{fig:Feynman-di-lepton}
\end{center}
\end{figure}

The right panel of Fig.\ref{fig:di-lepton} shows the result for
the muon-pair production, where the four-momenta dependent
production rate is given by 
\begin{eqnarray}
\frac{ d\Gamma }{ d^4 Q } 
= \frac{ -\alpha g^{\mu\nu} {\rm Im} \Pi_{\mu\nu}(Q) }
{ 12 \pi^4  Q^2 ( {\rm e}^{\beta q_0} - 1 ) }\cdot\big[1+\frac{2m_{\mu}^2}{q^2}]
\big[1-\frac{4m_{\mu}^2}{q^2}]^{1/2},
\label{eq:drate-mu}
\end{eqnarray}
instead of Eq.(\ref{eq:drate-0}).

One sees that some enhancement of the muon-pair production rate
over the normal one as in the case of the lepton-pair production.
However, the enhancement may  not be sufficient to detect by experiments.

\begin{figure}
\begin{center}
\includegraphics[scale=.4, angle=-90]{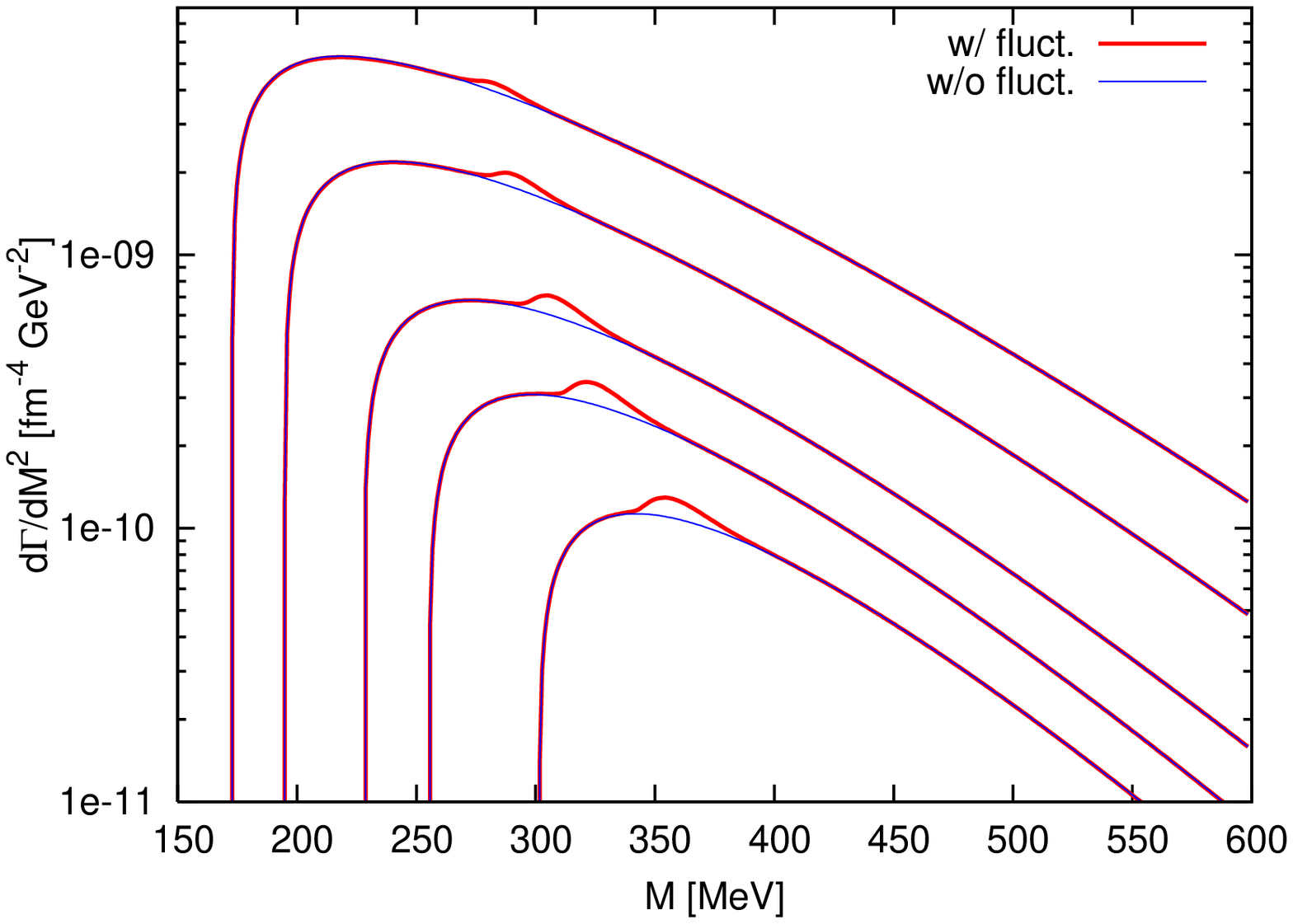}
\includegraphics[scale=.4, angle=-90]{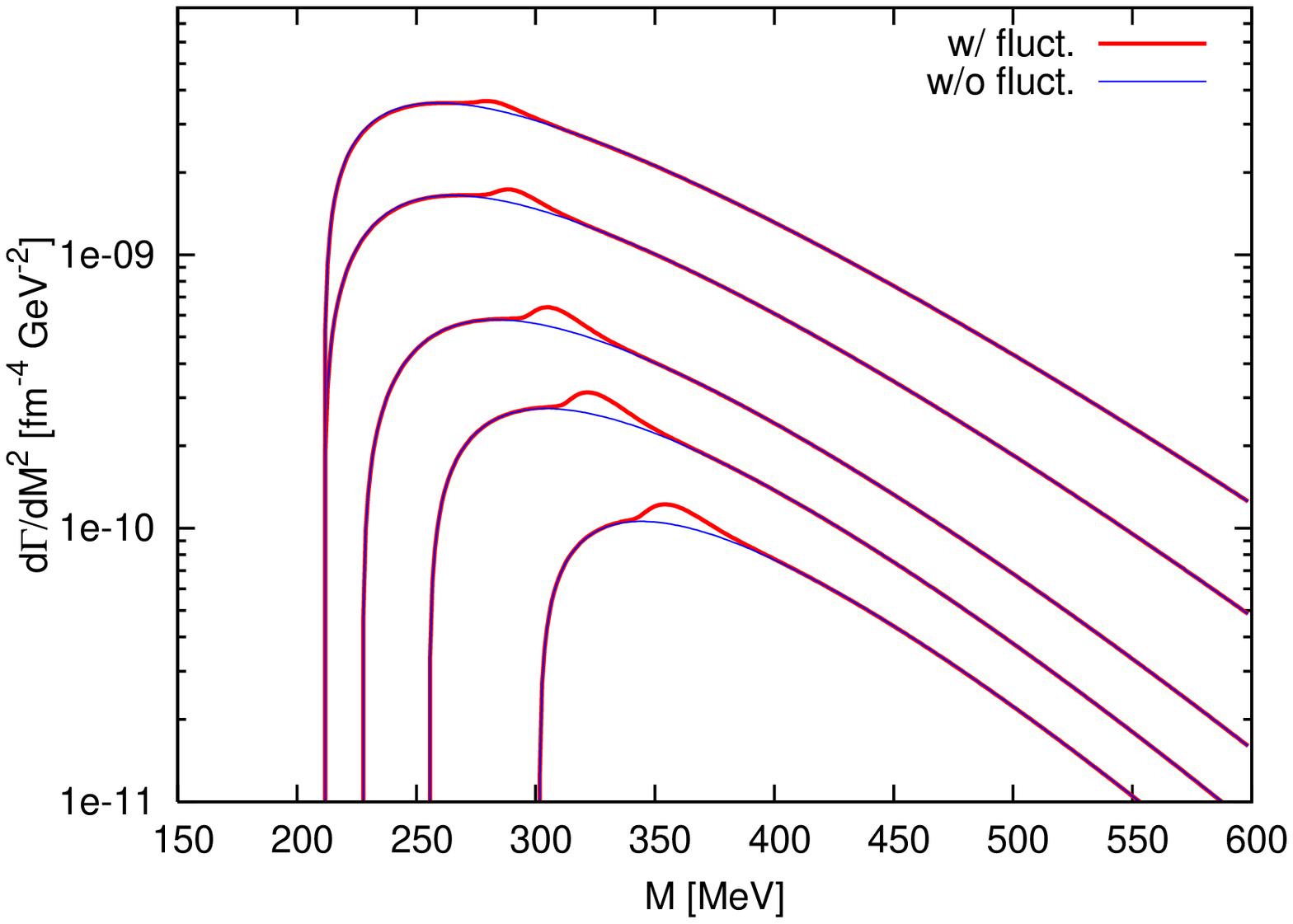}
\caption{
Left panel: the di-electron production rate for temperatures,
1.3$T_c$, 1.2$T_c$, 1.1$T_c$, 1.05$T_c$ and 1.01$T_c$; the upper curves
correspond to the higher temperatures.
Right panel: the di-muon production rate. Others are the same as the
left panel.
}
\label{fig:di-lepton}
\end{center}
\end{figure}

\section{Summary and concluding remarks}

We have first emphasized the importance of the notion of the soft modes
of QCD phase transitions at temperatures above $T_c$; they may be
hadronic excitations above $T_c$.
We have discussed the two QCD phase transitions, i.e., the 
color superconductivity and the chiral transition.
 The soft modes of the so called 2SC of the color superconductivity
can cause an enhancement of the electron-pair production in rather
 low invariant mass region.
 We have noticed that the soft mode around the critical end point 
where the transition is second order is a kind of 
density fluctuations coupled with the $\sigma$-meson like fluctuations. 
This soft modes have the dispersion relation which is space-like.
 Nevertheless, we explored how the sigma mesonic modes in the time-like 
region affect the di-lepton productions; the violation 
of the charge conjugation at finite
$\mu$ is responsible for the fact that the sigma mode can affect 
to the photon self-energy in the hot medium with the non-vanishing $\mu$.

{\bf Acknowledgements}\\ 
 T. K. is supported by a Grant-in-Aid
  for Scientific Research by Monbu-Kagakusyo of Japan
  (No.~17540250).
 M. K. is supported by a Grant-in-Aid
  for Scientific Research by Monbu-Kagakusyo of Japan
  (No.~19840037).
  Y. N. is supported by a JSPS Grant-in-Aid for Scientific
  Research (\#18740140).
  This work is supported by the Grant-in-Aid for the 21st Century COE 
  ``Center for Diversity and Universality in Physics" of Kyoto
  University.


\begin{thebibliography}{99}

%
\bibitem{Abuki:2005ms}
  H.~Abuki and T.~Kunihiro,
  Nucl.\ Phys.\  A {\bf 768} (2006) 118.
%
\bibitem{Kitazawa:2001ft}
  M.~Kitazawa, T.~Koide, T.~Kunihiro and Y.~Nemoto,
  Phys.\ Rev.\ D {\bf 65}, 091504 (2002).
%
\bibitem{Kitazawa:2003cs}
  M.~Kitazawa, T.~Koide, T.~Kunihiro and Y.~Nemoto,
  Phys.\ Rev.\ D {\bf 70}, 056003 (2004).
%
\bibitem{Kitazawa:2005vr}
  M.~Kitazawa, T.~Koide, T.~Kunihiro and Y.~Nemoto,
  Prog.\ Theor.\ Phys.\  {\bf 114}, 117 (2005).
%
\bibitem{Giannakis:2003am}
  I.~Giannakis and H.~c.~Ren,
  Nucl.\ Phys.\ B {\bf 669}, 462 (2003).
%
\bibitem{Matsuzaki:1999ww}
  M.~Matsuzaki,
  Phys.\ Rev.\ D {\bf 62}, 017501 (2000).
%
\bibitem{Abuki:2001be}
  H.~Abuki, T.~Hatsuda and K.~Itakura,
  Phys.\ Rev.\  D {\bf 65} (2002) 074014.
%
\bibitem{LL58}
  see, for example,
  L.D.~Landau and E.M.~Lifshiz, Statistical Physics 
  (Pergamon, New York, 1958).
%
\bibitem{Voskresensky:2004jp}
  D.N. Voskresensky, Phys. Rev. C {\bf 69}, 065209 (2004).
%
\bibitem{HTSC}
  Y.~Yanase, T.~Jujo, T.~Nomura, H.~Ikeda, T.~Hotta and K.~Yamada,
  Phys. Rep. {\bf 387}, 1 (2003).
%
\bibitem{TS99}
  As a experimental review,
  T.~Timusk and B.~Statt, Rep. Progr. Phys. {\bf 62}, 61 (1999).
%
\bibitem{Kitazawa:2005pp}
  M.~Kitazawa, T.~Kunihiro and Y.~Nemoto,
  Phys.\ Lett.\ B {\bf 631}, 157 (2005).
%
\bibitem{ref:AL-MT}
  L.G. Aslamazov and A.I. Larkin,
  Sov. Phys. Solid State {\bf 10}, 875 (1968);
  K. Maki, Prog. Theor. Phys. {\bf 40}, 193 (1968);
  R.S. Thompson, Phys. Rev. B{\bf 1}, 327 (1970).
%
\bibitem{LeBellac}
  See for example, M. Le Bellac, {\it Thermal Field Theory}
  (Cambridge University Press, Cambridge, England 1996).
%
\bibitem{KKprep}
  M.~Kitazawa and T.~Kunihiro, unpublished.
  A preliminary result is presented at
  the JPS meeting at Kochi University, Sep.,2004.
%
\bibitem{jai:2001} 
P.~Jaikumar, R.~Rapp and I.~Zahed, 
Phys.\ Rev.\ C {\bf 65}, 055205 (2002). 
%
\bibitem{HK84}
T.~Hatsuda and T.~Kunihiro, Phys. Lett. {\bf 145}, 7 (1984);
Prog. Theor. Phys. {\bf 75}, 765 (1985); Phys. Rev. Lett.
{\bf 55}, 158 (1985); 
Phys. Rep. {\bf 247}, 221 (1994).
%
\bibitem{karsch02} F. Karsch, Lect. Notes in Physics,
{\bf 583}, 209 (2002).
%
\bibitem{RHIC}
  M.~Gyulassy and L.~McLerran,
  Nucl.\ Phys.\  A {\bf 750} (2005) 30.
%
%
%
%
\bibitem{Lattice}
  M.~Asakawa and T.~Hatsuda,
  Phys.\ Rev.\ Lett.\  {\bf 92}, 012001 (2004);
%
  S.~Datta, F.~Karsch, P.~Petreczky and I.~Wetzorke,
  Phys.\ Rev.\ D {\bf 69}, 094507 (2004);
%
  T.~Umeda, K.~Nomura and H.~Matsufuru,
  Eur.\ Phys.\ J.\ C {\bf 39S1}, 9 (2005);
%
%
%
%
\bibitem{asakawa-yazaki}
M.~Asakawa and K.~Yazaki, Nucl. Phys. {\bf A504}, 668 (1989);\\
M. Lutz, S. Klimt and W. Weise, Phys. Lett. {\bf B249}, 386 (1990);
Nucl. Phys. {\bf A542}, 521 (1992).
%
\bibitem{TK89}
T. Kunihiro, Phys. Lett. {\bf B 219}, 363 (1989); {\bf B245}, 687 (1990)(E);
Nucl. Phys. {\bf B351}, 593 (1991).
%
\bibitem{stephanov}
M. Stephanov,
 Prog. Theor. Phys. Suppl.{\bf 153},139 (2004).
%
\bibitem{kitazawa02}
M.Kitazawa, T. Koide, T. Kunihiro and Y. Nemoto, Prog. Theor. Phys.
{\bf 108}, 929 (2002).
%
\bibitem{Hatsuda-anomaly}
  T.~Hatsuda, M.~Tachibana, N.~Yamamoto and G.~Baym,
  Phys.\ Rev.\ Lett.\  {\bf 97} (2006) 122001.
  
%
\bibitem{fujii}
 H.~Fujii,
  Phys.\ Rev.\   {\bf D67} (2003) 094018;
H. Fujii and M. Ohtani, Phys. Rev. {\bf D70} (2004) 014016.
%
\bibitem{son-stephanov}
D. T. Son and M. Stephanov, Phys. Rev. {\bf D70 } (2004)056001.
%
\bibitem{nemoto}
Y. Nemoto, M. Kitazawa and T. Kunihiro, in preparation.
%
\bibitem{baryon}
T. Kunihiro, Phys. Lett. {\bf B271}, 395 (1991).
%
\end{thebibliography}
\end{document}